\documentclass[pre,
 twocolumn,
eqsecnum,
amsmath,amssymb,
amsmath,amssymb, aps,
]{revtex4}

\usepackage{epsfig}
\usepackage{amsfonts}
\usepackage{amsmath}
\usepackage{amssymb}
\usepackage{bm}
\usepackage[framemethod=PStricks]{mdframed}
\usepackage{lipsum}
\usepackage{float}
\usepackage{color}
\usepackage{pdflscape}
\usepackage{graphicx}
\usepackage{todonotes}

\newcommand{\be}{\begin{equation}}
\newcommand{\ee}{\end{equation}} 
\newcommand{\bea}{\begin{eqnarray}} 
\newcommand{\eea}{\end{eqnarray}}

\begin{document}

\title{Toy Model for Vortex Ring-Assisted Particle Drag in Superfluid Counterflow}

\author{L. Moriconi}
\affiliation{Instituto de F\'\i sica, Universidade Federal do Rio de Janeiro, \\
C.P. 68528, CEP: 21945-970, Rio de Janeiro, RJ, Brazil}


\begin{abstract}
The interpretation of data obtained from particle image/tracking velocimetry in the study of superfluid flows has been so far a challenging task. Tracking particles (as solid hydrogen or deuterium) are attracted to the cores of quantized vortices, so that their dynamics can be strongly affected by the surrounding vortex tangle. Previous phenomenological arguments indicate that tracking particles and micro-sized vortex rings could form bound states (denoted here as VRP states). While a comprehensive description of the vortex ring-particle bonding mechanism has to deal with somewhat involved flow configurations, we introduce a simplified two-dimensional model of VRP states, which captures essential qualitative features of their three-dimensional counterparts. Besides an account of known experimental and numerical observations, the model proves to be of great heuristic interest. In particular, it sheds light on the important role played by viscous dissipation (due to the normal component of the fluid), the Magnus force, and topologically-excited vortex rings in the stability and dynamics of VRP states.
\end{abstract}


\maketitle

\section{Introduction}


A vigorous boost of interest in quantum turbulence has been witnessed along the last two decades, with significant progress being made both on the numerical and experimental research fronts \cite{skr_etal,vinen_niemela,ara_tsu_nemi,volo,halp_tsu,vinen,skr,bradley_etal,BarSkrSree,tsu,BarSerBagg,bar_par,tsu_fuji_yui,cidrim,galan_etal}. While ingenious and landmark experiments established long ago the existence of quantum vortices \cite{vinen58,whit_zimm} -- the main actor of quantum turbulence -- it has not been until more recent times that optical techniques originally applied to classical flows, like particle image or particle tracking velocimetry (PIV or PTV) \cite{raffel_etal}, made their debut in the field of quantum fluid dynamics \cite{zhang,bewley_etal,chago_sci,guoa}, allowing the efficient visualization of these topologically stable flow structures as they evolve in time.

\begin{figure}[ht]
\hspace{0.0cm} \includegraphics[width=0.25\textwidth]{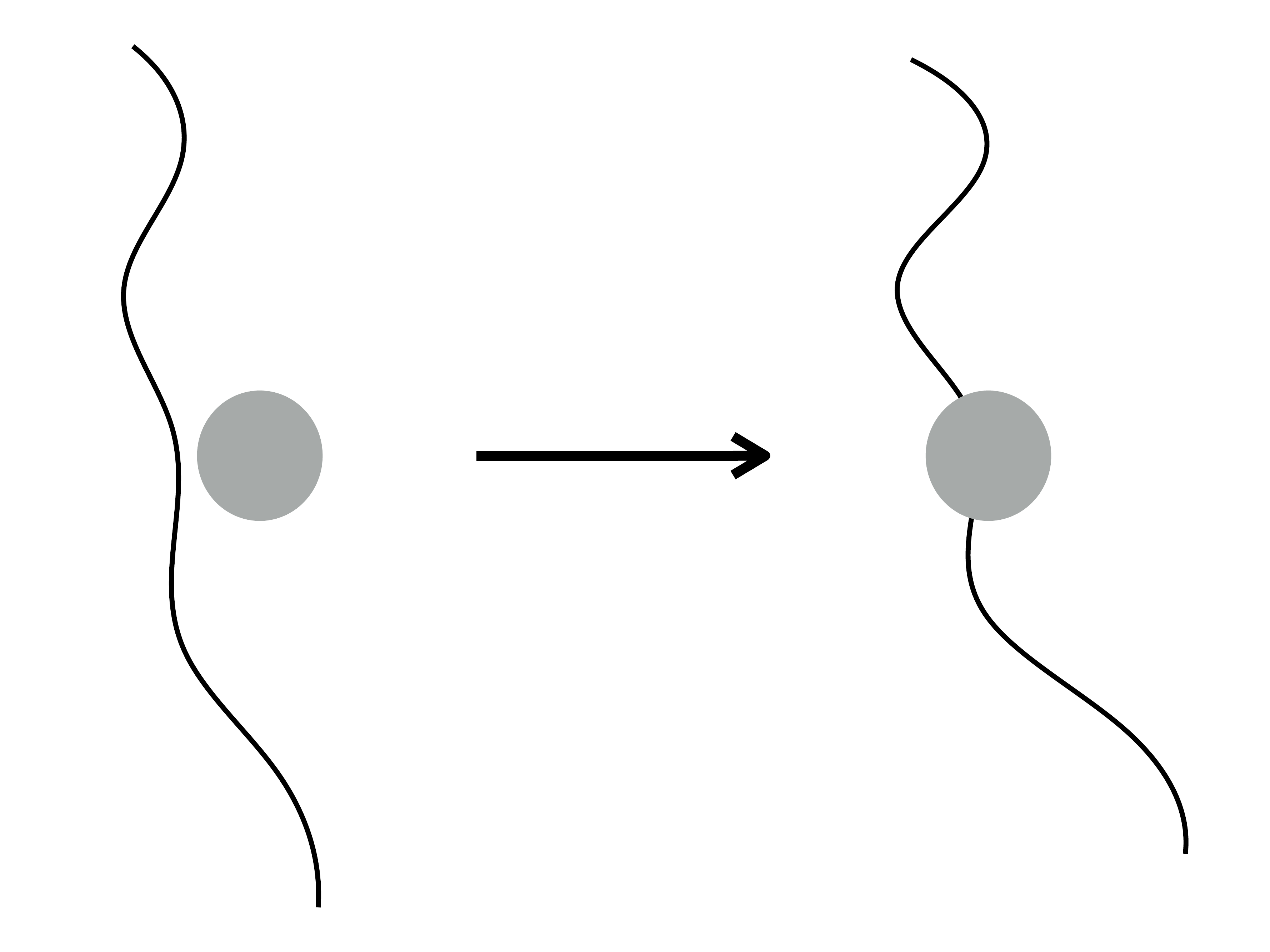}
\vspace{0.0cm}
\caption{A particle can get trapped to the core of a quantized vortex filament, due to the strong pressure
depletion around it. When this happens, the particle's trajectory will be essentially ruled by the evolution of the 
host vortex filament.}
\label{vcp}
\end{figure}

The application of PIV/PTV techniques to superfluid helium has been implemented in general with micro-sized particles produced from the {\it{in situ}} solidification of injected gaseous hydrogen or deuterium \cite{bewley_etal,chago_sci,guoa}. Some of these particles are ``swallowed" by quantum vortices, as it is illustrated in {\hbox{Fig. \ref{vcp}}}. As the superfluid vortices move, they carry with them the attached particles, which, then, work as vortex tracers. The tight attachment of particles to vortex cores is due to the fact that even small displacements of the vortices lead to strong pressure gradients that overcome by a large extent buoyancy effects and the viscous drag exerted on the particles by the normal component of the flow. This simple phenomenological background has paved the way for a wave of remarkable experiments, as the visualization of quantum vortex reconnection \cite{bewley_etal2}, the decay of quantum vortex rings \cite{bewley_sree}, the dynamical breaking of superflow regimes \cite{chago_sci2}, and even the observation of Kelvin waves on quantum vortices \cite{fonda_etal}.

In flow regimes where vortex filaments are sparse enough, suitable for the application of PIV/PTV in the study of local flow properties, the vast majority of tracking particles do not spend most of their times confined to vortex cores. Instead, they wander through the vortex tangle, affected in many different ways by its related intermittent pressure field. It is clear, therefore, that models for the interaction between particles and quantum vortices are absolutely in order to interpret the data provided by PIV and PTV methodologies in the experimental study of superfluid flows. 

It has been found, very peculiarly, from PTV analyses of counterflow turbulence, that the velocities of tracking particles have, depending on the level of vortex filament length density, peaked distributions centered around $v_n$ or both $v_n$ and $v_n/2$ \cite{zhang_sci,mastracci_guo}, where $v_n$ is the local velocity of the normal component of the flow. Even more oddly, a minor fraction of the particles, which are likely not to be bound to vortex cores, have been found to move opposite to the direction of the normal flow. A theoretical account of the $v_n/2$ velocity peak has been given by Sergeev et al. \cite{serg_bar_kivo}, from the assumption that tracking particles could form bound states with vortex rings. In this connection, it has been known that collisions between quantum vortices and particles can produce small vortex rings \cite{kivo_serg_bar,kivo_wilkin}, a possible initial stage for the formation of such states. 

Our aim in this work is to dig further into the still barely explored coupled dynamics of conjectured vortex ring-particle bound states (VRP states, for short) by means of an essentially heuristic approach. Restricting our attention to axisymmetric flow configurations, we are able to introduce a two-dimensional version of VRP states, which leaves aside complicating three-dimensional quantitative details and preserves important qualitative features, as the topology of the superfluid flow around particles, Stokes drag, the existence of Magnus forces, etc. The good news is that one can greatly benefit from the much simpler analytical structure of two-dimensional incompressible and inviscid flows. As we will see, it is possible to address in this way phenomenologically interesting facts, a number of them yet unnoticed in the literature. It is worth emphasizing that two-dimensional toy models have been studied as useful sources of insight in the superfluid dynamics of particle-vortex interactions \cite{kivo_serg_bar,wang_etal}.

This paper is organized as follows. We discuss, in Sec. II, the formation of stable VRP states in superfluid counterflow regimes from a phenomenological perspective. We, then, address the formulation of two-dimensional toy models as heuristic tools to capture, in a qualitative way, relevant aspects of the analogous three-dimensional flows. The dynamical model for the evolution of VRP states is detailed in Sec. III. Next, in Sec. IV, we classify the asymptotically stable flow regimes of the vortex ring-particle system (not necessarily in the form of VRP states) and discuss how to explore the relatively huge eight-dimensional phase-space which underlies the proposed dynamical model. In Sec. V, we obtain phase diagrams that bring information on the resulting asymptotic flow regimes as the modeling parameters are varied. In Sec. VI, we examine the dynamical balance of velocity contributions in stable VRP states, corroborating ideas presented in Sec. II. Finally, in Sec. VII, we summarize our findings and point out directions of further research.

\section{Phenomenological Ingredients}

Our analysis is inspired by superfluid counterflow regimes similar to the ones investigated in experiments \cite{zhang_sci,mastracci_guo,duda} and numerical simulations \cite{kivo_serg_bar,kivo_wilkin}. We mean approximately equal mass densities $\rho_n$ and $\rho_s$ for the normal and superfluid components of the condensate, respectively, a condition realized at temperatures close to $2 \hbox{ K}$, and spherical tracking particles with radii around $2.5 \hbox{ $\mu$m}$.  The counterflow velocities are assumed to be bounded from above by $10 \hbox{ mm/s}$, which ensures, from Vinen's law for the mean density of vortex filament length \cite{vinenL1,vinenL2}, that the mean distance $\delta$ between vortices is much larger than the particles' dimensions ($\delta > 50 \hbox{ $\mu$m}$).

\begin{figure}[ht]
\hspace{0.0cm} \includegraphics[width=0.45\textwidth]{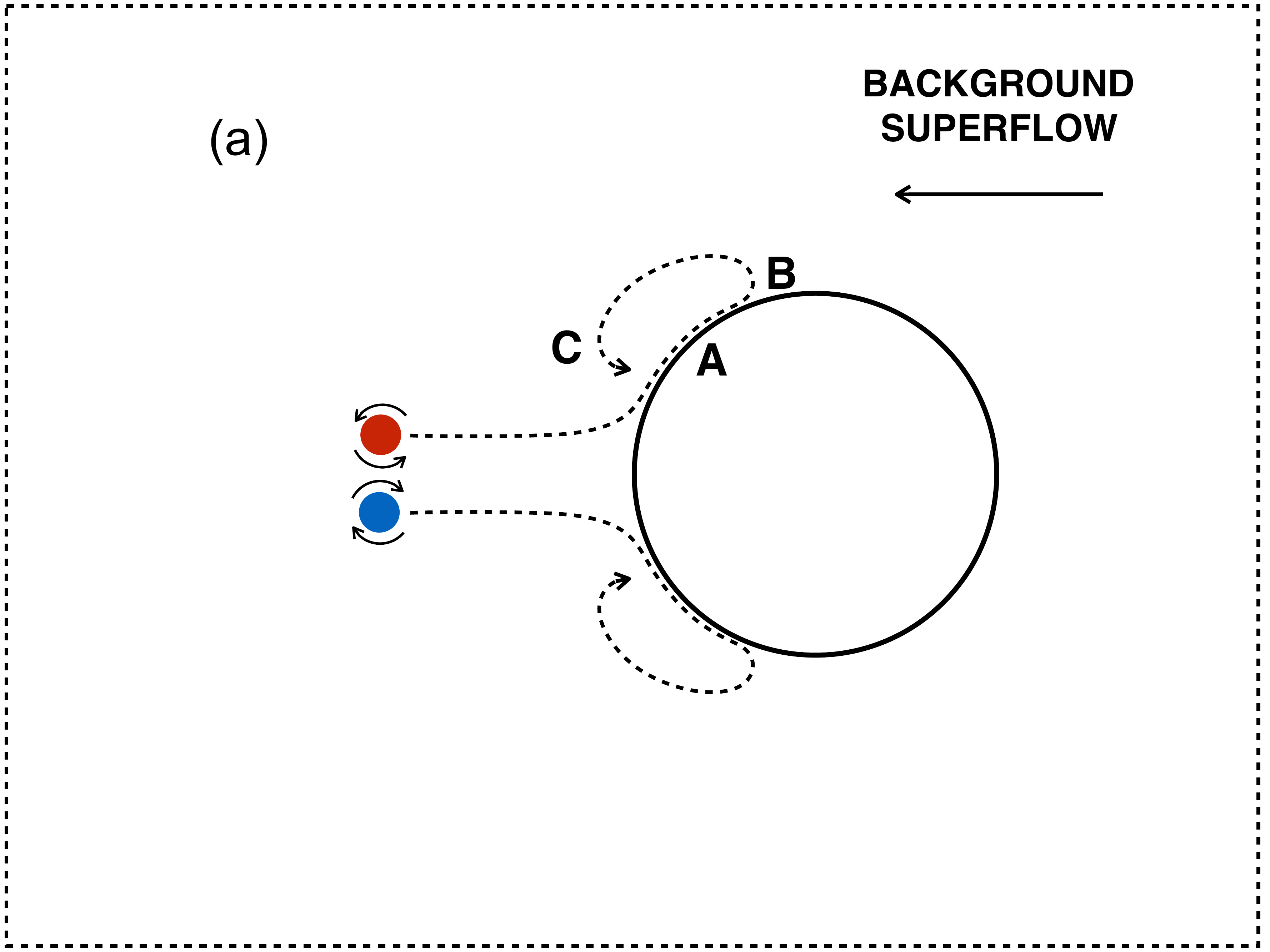}
\hspace{0.0cm} \includegraphics[width=0.45\textwidth]{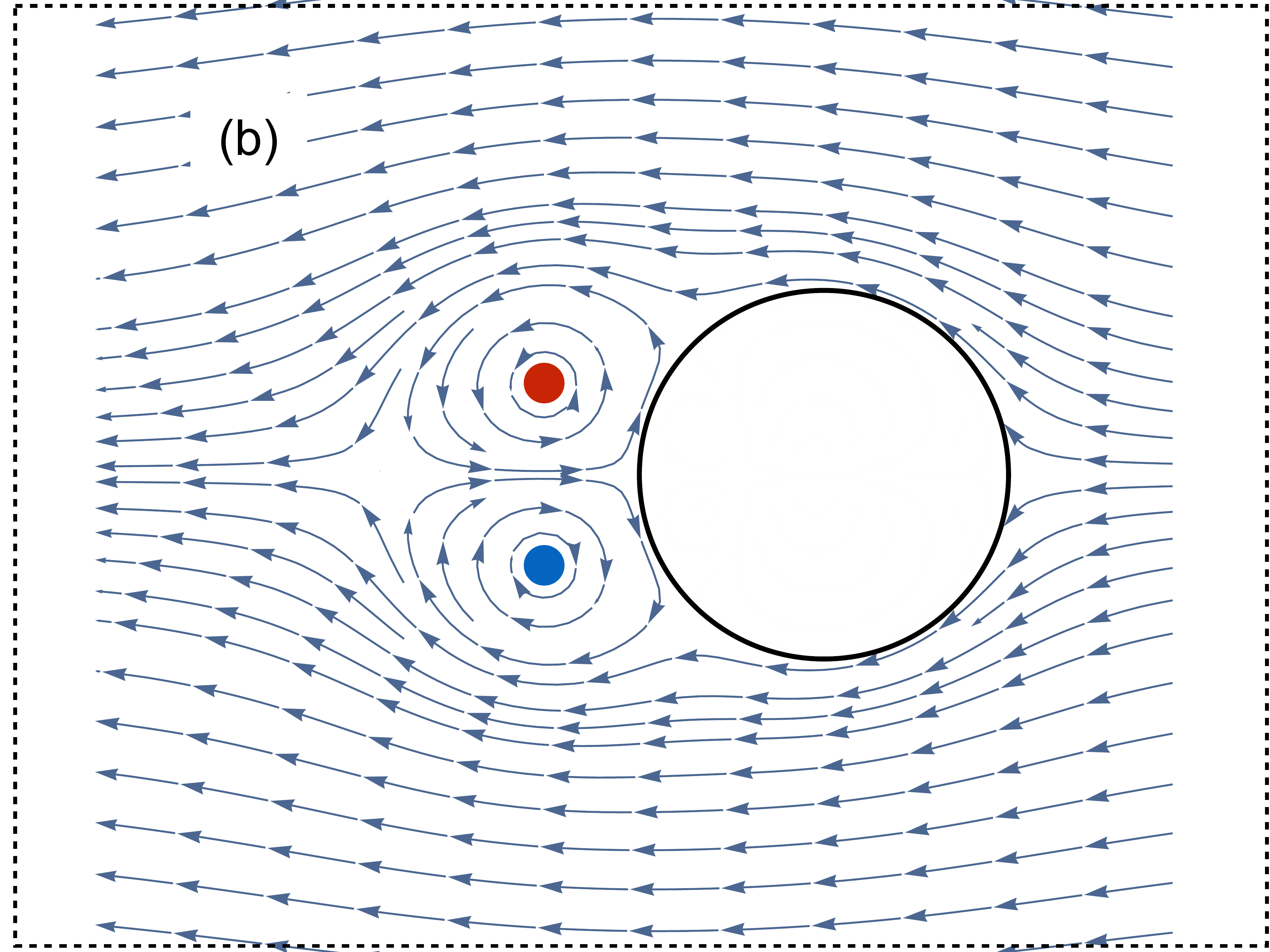}
\caption{(a) Sketch of the formation of VRP states as a three-stage process A$\rightarrow$B$\rightarrow$C. The vortex ring evolution is represented on a symmetry plane cut, as seen from the particle's reference frame (see the text, Sec. II, for details); (b) a vortex dipole hydrodynamically interacts with a solid disk in the presence of a two-dimensional superfluid stream.}
\label{portrait}
\end{figure}

In order to investigate the dynamics and stability of VRP states, we note that an interesting idealized situation is given by the case where the axial symmetry axis of a vortex ring is postulated to be parallel to an asymptotically uniform background superflow and, furthermore, to cross the center of a similarly sized near particle. A symmetry plane cut of this axisymmetric configuration is shown in {\hbox{Fig. \ref{portrait}a}}. Notice, from the picture, that the circulation of vorticity carried by the singular vortex ring is oriented (using the right-hand rule) opposite to the superflow direction (which, equivalently, means that the vortex ring is oriented parallel to the normal flow).

The mechanism for the formation of stable VRP states, as observed from the particle's reference frame, is schematized in {\hbox{Fig. \ref{portrait}a}}, as a three-stage process A$\rightarrow$B$\rightarrow$C. During stage A, the vortex ring gets advected by its image vortex ring field towards regions where the background superflow velocity is intensified. The vortex ring is then advected backwards by the locally stronger background superflow at stage B. As the vortex ring moves away from the particle, the superposition of its self-induced flow and the flow produced by its image dominates the dynamics along the axial direction, stage C, leading, after consecutive loops, to an asymptotically stable limit cycle. A bit of reflection tells us that VRP states cannot be stabilized if the vortex ring orientation or the superfluid background flow, or even both, are reversed in {\hbox{Fig. \ref{portrait}a}}. The discussion on the stabilization of VRP states will be retaken in Sec. VI, with further quantitative elements.

Our heuristic strategy in this work is to avoid computationally costing analyses, and center attention on the main qualitative properties of VRP states. As it is clearly suggested from the symmetry plane cut provided in {\hbox{Fig. \ref{portrait}a}}, we can devise a two-dimensional model that mimics the dynamics of VRP states, where three-dimensional spherical particles give place to disk-shaped particles, and vortex rings are replaced by planar vortex dipoles. {\hbox{Fig. \ref{portrait}a}}, thus, is to be taken, from now on, as a visual reference for two-dimensional modeling. Following this line of thought, {\hbox{Fig. \ref{portrait}b}} depicts exact two-dimensional superfluid streamlines, which have the same topology as the ones obtained from a symmetry plane cut of the three-dimensional superflow associated to an axisymmetric VRP state. 

\section{Two-Dimensional Toy Model}

Relying on the usual complex representation for positions and velocities in two-dimensional space, let, in the $xy$-plane, the center of a tracking particle which has mass $M$ and radius $R$ have, at time $t$, position $z_p(t)$ and velocity $u(t) = \dot z_p(t)$. Quantum vortices which carry circulations $\phi$ and $- \phi$ have, respectively, positions $z_+(t)$ and $z_-(t)$. The far background velocities of the normal and the inviscid components of the flow have fixed velocities $V$ and $-V$, both real numbers. This is just a counterflow regime where the mass densities of the normal and inviscid components are the same. 

We are interested to study mirror symmetric flow configurations where $z_p(t)$ is real and $z_+(t) = z^*_-(t)$, with initial conditions Re$[z_p(t=0)]> 
{\hbox{Re}}[z_\pm(t=0)]$ and Im$[z_+(t=0)] > 0$. Abstracting coordinate axes, this is the situation shown in {\hbox{Fig. \ref{portrait}b}}, for the case where the particle is momentarily at rest, that is, $u=0$.

As it is discussed in standard text-books \cite{acheson}, incompressible and inviscid two-dimensional flows can be completely described in terms of complex potentials. Even though this is a fundamental technical point in our analysis, it is not the whole story, since we want to bring into two-dimensional modeling at least two important phenomenological aspects of three-dimensional dynamics, namely, 
\vspace{0.2cm}

(i) particles are subject to viscous forces given by Stokes' law, viz.,
\be
F_D = CR(V - u) \ , \ \label{dragforce}
\ee
where $C$ plays the role of a friction coefficient, and 
\vspace{0.2cm}

(ii) the evolution of quantum vortices is determined by Magnus and drag contributions, besides background advection, in close analogy with the three-dimensional Schwarz's description of vortex filament dynamics \cite{schwarz}.
\vspace{0.2cm}

\noindent To implement (ii), we note that a straightforward adaptation of Schwarz's vortex filament dynamics
to planar positive/negative vortices which move with velocities $v_\pm$ and carry circulations $\pm \phi$ 
is given by
\be
v_\pm = v_s^\pm \pm i \alpha (v_n^\pm - v_s^\pm) + \alpha' (v_n^\pm - v_s^\pm) \ , \  \label{vpm1}
\ee
where $v_n^\pm$ and $v_s^\pm$ are the normal and superfluid velocities at the positions of the positive 
and negative vortices, and $\alpha$ and $\alpha'$ are the usual mutual friction coefficients that model 
the coupling between the normal fluid and quantum vortices \cite{donnelly,bar-donn}. 

Eq. (\ref{vpm1}) could be used as it stands, with the help of postulated analytic expressions for 
$v_n^\pm$ and $v_s^\pm$. However, to attain a simpler formulation for the cases of interest (VRP states),
$v_n^\pm$ can be neglected due to the fact that the 
vortices will stay close to the particle's surface where the normal fluid velocity is suppressed, 
whereas the superfluid velocity never vanishes and is most of the time approximately the same as the 
one of the far background superflow, that is, $- V$. Taking these considerations into account, we 
write down, more simply, that
\be
v_\pm = v_s^\pm \pm i \alpha V + \alpha' V  \ . \ \label{vpm2}
\ee
In other words, in the computation of vortex velocities, the interaction between the normal fluid and quantum vortices is encoded in the form of constant shifts of velocity, $\pm i \alpha V + \alpha' V$, over the fluctuating superfluid velocity background $v_s^\pm$. The rationale for not substituting the first term on the RHS of (\ref{vpm2}) by $V$ is that the coefficients $\alpha$ and $\alpha'$ are in general small numbers (the second and third terms on the RHS of (\ref{vpm1}) are, as a matter of fact, small perturbative corrections to the background superflow velocities $v_s^\pm$).

Let $W(z)$ be the complex velocity potential defined at positions which are free from vortex singularities. To define the velocity fields at the vortex positions, we introduce the complex velocity potentials $\hat W_+(z)$ and $\hat W_-(z)$, which are just $W(z)$ subtracted by the singular contributions associated to the respective self-induced velocity fields, that is,
\bea
&& \hat W_+(z) \equiv W(z) + i \frac{\phi}{2 \pi} \ln(z-z_+) \ , \ \label{Wp} \\
&& \hat W_-(z) \equiv W(z) - i \frac{\phi}{2 \pi} \ln(z-z_-) \ , \ \label{Wm}
\eea
so that
\be
(v_s^\pm)^* = \left. \frac{d}{d z} \hat W_\pm(z) \right |_{z = z_\pm} \label{vpmcc} \ . \
\ee
Observing, now, that $v_\pm = \dot z_ \pm$, we take Eqs. (\ref{dragforce}) and (\ref{vpm2}--\ref{vpmcc}) to establish a closed set of evolution equations for the vortex dipole-particle dynamical system, 
\bea
&& \left. \dot z_+^* = V (\alpha' - i \alpha -1) + \frac{d}{d z} \hat W_+(z) \right |_{z = z_+} \ , \ \label{zpdot}  \\
&& \left. \dot z_-^* = V ( \alpha' + i \alpha  -1)  +  \frac{d}{d z} \hat W_-(z) \right |_{z = z_-} \ , \ \label{zmdot} \\
&& \dot z_p = u \ , \  \label{zpartdot} \\ 
&& M \dot u^* =  CR(V - u^*) + i \frac{\rho_s}{2} \oint_{\partial D} \left ( \frac{dW}{dz} \right )^2 dz   \ . \  \label{mudot} \nonumber \\ 
\eea
We have, above,
\be
W(z) = W_+(z) + W_-(z) + W_p(z) \ , \ \label{Wz}
\ee
where
\bea
&& W_+(z) = \nonumber \\
&&= i \frac{\phi}{2 \pi}  \ln \left [ \frac{(z-z_p)(z_+^* - z_p^*) - R^2}{z-z_+} \right ]  \ , \  \\
&& W_-(z) = \nonumber \\
&& = - i \frac{\phi}{2 \pi} \ln \left [ \frac{(z-z_p)(z_-^* - z_p^*) - R^2}{z-z_-} \right ] \ , \ 
\eea
are the complex potentials produced by the positive and negative vortices together with their respective vortex images, and 
\be
W_p(z) = - \frac{R^2(V + u)}{z-z_p}
\ee
is the complex potential for the flow around the circular particle in its reference frame. The second term on the RHS of Eq. (\ref{mudot}) is the pressure force on the particle, as prescribed by Blasius theorem \cite{acheson}. The closed contour integral is performed along the counterclockwise oriented boundary $\partial D$ of the solid particle.

It should be clear, furthermore, that we are not worried with added mass issues in the equations of motion \cite{acheson,falko}, as far as this is a point of importance only in detailed quantitative analyses. For all purposes, the mass parameter $M$ in Eq. (\ref{mudot}) is to be taken as an effective quantity.

Besides the two dimensionless parameters $\alpha$ and $\alpha'$, the dynamical system (\ref{zpdot}--\ref{mudot}) is, in principle, defined from the six additional dimensionful parameters $V, C, M, \phi, \rho_s$, and $R$. We can define units of mass, length, and time from the values of $V$, $R$, and $M$. Our dynamical system can then be recast in dimensionless form through the substitutions
\bea
&& u \rightarrow  V u \ , \ t \rightarrow \frac{R}{V} t \ , \  z \rightarrow  Rz \ , \  \nonumber \\
&& z_+ \rightarrow  Rz_+ \ , \ z_- \rightarrow  Rz_- \ , \ z_p \rightarrow  Rz_p   \ , \ \label{prmts}
\eea
and
\be
 C \equiv \frac{MV}{R^2} a \ , \ 
\rho \equiv \frac{M}{R^2} b \ , \ 
\phi \equiv  VRc \ . \ \label{abc}
\ee
The above set of relations, (\ref{prmts}) and (\ref{abc}), allows us to rewrite Eqs. (\ref{zpdot}--\ref{mudot})
as
\bea
&& \left. \dot z_+^* =  \alpha' - i \alpha  - 1 + \frac{d}{d z} \hat W_+(z) \right |_{z = z_+} \ , \  \label{zzpdot} \\
&& \left. \dot z_-^* =  \alpha' + i \alpha - 1  +  \frac{d}{d z} \hat W_-(z) \right |_{z = z_-} \ , \ \label{zzmdot} \\
&& \dot z_p = u \ , \  \label{zzpartdot} \\ 
&& \dot u^* = a(1  - u^*) + i \frac{b}{2} \oint_{\partial D} \left ( \frac{dW}{dz} \right )^2 dz   \ , \  \label{udot}
\eea
where, now,
\bea
&& \hat W_+(z) = W(z) + i \frac{c}{2 \pi} \ln(z-z_+) \ , \ \label{hatWp} \\
&& \hat W_-(z) = W(z) - i \frac{c}{2 \pi} \ln(z-z_-) \ , \  \label{hatWm}
\eea
with Eq. (\ref{Wz}) still holding, provided that $W_+(z)$, $W_-(z)$, and $W_p(z)$ be 
redefined as
\bea
&& W_+(z) = \nonumber \\
&& = i \frac{c}{2 \pi} \ln \left [ \frac{(z-z_p)(z_+^* - z_p^*) - 1}{z - z_+} \right ] \ , \ \label{Wp} \\
&& W_-(z) =   \nonumber \\
&& = - i \frac{c}{2 \pi} \ln \left [ \frac{(z-z_p)(z_-^* - z_p^*) - 1}{z-z_-} \right ] \ , \ \label{Wm}
\eea
and
\be
W_p(z) = - \frac{1 + u}{z-z_p} \ . \ \label{Wpart}
\ee
We find, therefore, that the vortex dipole-particle dynamical system is actually determined by five independent dimensionless parameters, i.e., $\alpha$, $\alpha'$, $a$, $b$, and $c$, where the last three ones are defined from (\ref{abc}). 

The exact evaluation of the complex integral in (\ref{udot}) yields
\bea
&&\oint_{\partial D} \left ( \frac{dW}{dz} \right )^2 dz =  
2i \frac{c^2}{(2 \pi)^3} \frac{z_+^* - z_-^*}{|z_+ - z_p|^2 - 1} + \nonumber \\
&& - 2i \frac{c^2}{(2 \pi)^3} \frac{z_+^* - z_-^*}{|z_- - z_p|^2 - 1} + \nonumber \\
&&+ \frac{c}{2 \pi^2} (1+u) \frac{(z_- - z_+)(z_+ + z_- - 2z_p)}{(z_p-z_+)^2(z_p-z_-)^2} \ . \ \label{p_int}
\eea
Since we focus our attention on flow configurations which are mirror symmetric with respect to the $x$-axis, 
it is not difficult to show that the first two terms on the RHS of Eq. (\ref{p_int}) cancel each other.

Our task, now, is to explore solutions of the coupled Eqs. (\ref{zzpdot}--\ref{udot}),
to verify if they can have in fact the form of VRP states. 

\section{Flow Regimes}

A rather elementary inspection of Eqs. (\ref{zzpdot}--\ref{udot}) shows that there are just two classes of flow regimes for the vortex dipole-particle dynamical system:
\vspace{0.2cm}

(i) In the {\it{particle-vortex decoupled}} flow regime, the particle decouples from the vortex dipole and even arbitrarily small Magnus contributions will make the distance between vortices to grow without bound, so that they become independent in the long run. 
The solid particle, on its turn, is dragged by the background normal flow until it reaches its limiting velocity.  Using original units, at asymptotically large times the particle's velocity approaches $V$, while the vortices' velocities tend to $V(\alpha' -1 \pm i \alpha)$.
\vspace{0.2cm}

(ii) In the regimes of {\it{forward}} or {\it{backward-moving}} VRP states, the distance between vortices and the particle is upper bounded, and all of them, 
respectively, move, in the mean, along or opposite to the direction of the background normal flow.
\vspace{0.2cm}

In order to investigate how flow regimes are distributed over solutions of Eqs. (\ref{zzpdot}--\ref{udot}), it is necessary to define a set of initial conditions $z_+(0)=z^*_-(0)$ (taking, without loss of generality $z_p(0)=0$) and $u(0)$, besides values for the dynamical system parameters $a$, $b$, and $c$, introduced in (\ref{abc}), and the mutual friction  coefficients $\alpha$ and $\alpha'$. We are, therefore, interested to study the eight-dimensional phase-space $\Lambda$, a subset of $\mathbb{R}^8$, defined by the coordinates $\left ( {\hbox{Re}}[z_+(0)], {\hbox{Im}}[z_+(0)], u(0),\alpha, \alpha',a,b,c \right )$. We want to partition the points of $\Lambda$ into equivalence classes labelled by asymptotic flow regimes.

It is reasonable to assume that the production of a VRP state is more likely if the particle and the vortex dipole are initially as close as possible, with the former at rest. For that reason, we take, in all of our simulations,
\bea
&&z_+(0)=z^*_-(0)= -1+\frac{1}{2}i \ , \ \label{z0} \\
&&u(0)=0 \ . \ \label{u0}
\eea
We still need to select physically interesting regions in the five-dimensional cut of $\Lambda$ spanned by the $a,b$,$c$, $\alpha$ and $\alpha'$ parameters, with fixed $z_+(0)$ and $u(0)$. It would be, in fact, very difficult to figure out such regions without relying on physical inputs. For the sake of parameter estimation, we build a bridge between two-dimensional and three-dimensional superfluid phenomenologies, resorting on dimensional arguments.

Let $\rho_p$, $\nu$ and $\kappa$ be, respectively, the particle mass density, the kinematic viscosity of the normal component of the fluid, and the circulation quantum. We may readily put forward the estimates
\bea
&& \rho \sim \rho_s R \ , \ \label{est1} \\
&& M \sim \rho_p R^3 \ , \ \label{est2} \\
&& C \sim \rho_n \nu \ , \ \label{est3} \\
&& \phi \sim \kappa \ , \ \label{est4}
\eea
where the quantities along the left and right hand sides of (\ref{est1}-\ref{est4}) refer, respectively, to 
two-dimensional and three-dimensional contexts. Therefore, it follows, from (\ref{abc}) and (\ref{est1}-\ref{est4}), that
\bea
&& a = \frac{R^2}{MV} C \sim \frac{\rho_n \nu }{\rho_p V R}  \ , \ \label{aa} \\
&& b = \frac{R^2}{M} \rho \sim \frac{\rho_s}{\rho_p}  \ , \ \label{bb} \\
&& c = \frac{1}{VR} \phi \sim \frac{1}{VR} \kappa \ . \ \label{cc}
\eea
For temperatures around $T \simeq {\hbox{$2$ K}}$ as set in usual experiments, we have \cite{bar-donn} $\alpha \simeq 0.3$, $\alpha' \simeq 0.01$, ${\hbox{$\nu \simeq 2.6 \times 10^{-8}$ m$^2$/s}}$, and $\rho_n \simeq \rho_s \simeq 60$ Kg/m$^3$. We take, from reported experimental data \cite{duda}, $\rho_p \simeq \rho = \rho_s + \rho_n$, $V \simeq {\hbox{$10^{-2}$ m/s}}$, and $R \simeq 5 \times 10^{-6}$ m. Since ${\hbox{$\kappa \simeq 10^{-7}$ m$^2$/s}}$, we find, using (\ref{aa}-\ref{cc}),
\bea
&& a \sim \frac{\rho_n \nu }{\rho_p V R}  \simeq 0.3 \ , \ \label{anum} \\
&& b \sim \frac{\rho_s}{\rho_p} \simeq  0.5 \ , \ \label{bnum} \\
&& c \sim \frac{1}{VR} \kappa \simeq 2 \ . \ \label{cnum}
\eea

\section{Phase Diagrams}

The numerical values listed in (\ref{anum}-\ref{cnum}) should not be taken as strict definitions, but just as hints about phase-space regions where one should search for non-trivial flow regimes (that is, VRP states). We have investigated a set of four two-dimensional slices of $\Lambda$, defined as the regions $0 \leq a \leq 0.2 $ and $0 \leq c \leq 1$, for $b=0.5,1.0,2.0,$ and $4.0$, with fixed mutual friction  coefficients $\alpha =0.2$ and $\alpha'=0$.

Eqs. (\ref{zzpdot}-\ref{udot}) have been solved with the initial conditions (\ref{z0}) and (\ref{u0}) through the standard Euler's method
with time step $\epsilon = 10^{-4}$, for the total evolution time $T=10^2$. The values of parameters $a$ and $c$ have been independently incremented in steps of size $\delta a = 2 \times 10^{-3}$ and $\delta c = 10^{-2}$, respectively. Each one of the four phase diagrams shown in {\hbox{Fig. \ref{pd}}} is, thus, the result of $10^4$ complete simulation runs.

These phase diagrams show, notably, the occurrence of all of the three possible asymptotic flow regimes for the vortex dipole-particle dynamical system. Some observations are in order:
\vspace{0.2cm}

(i) The larger is $b$, the larger is the phase-space region associated to backward-moving VRP states. As it is clear from (\ref{udot}), $b$ defines the strength of pressure forces. This means that for larger $b$, the particle is more strongly attracted towards the vortex dipole and can be, in this way, carried (against viscous drag) by the vortex dipole during outer excursions of the limit orbits (trajectory sector that goes from $B$ to $C$ in Fig. \ref{portrait}a).

\begin{widetext}
\hspace{5.0cm}
\begin{figure}[b]  
\includegraphics[width=0.98\textwidth]{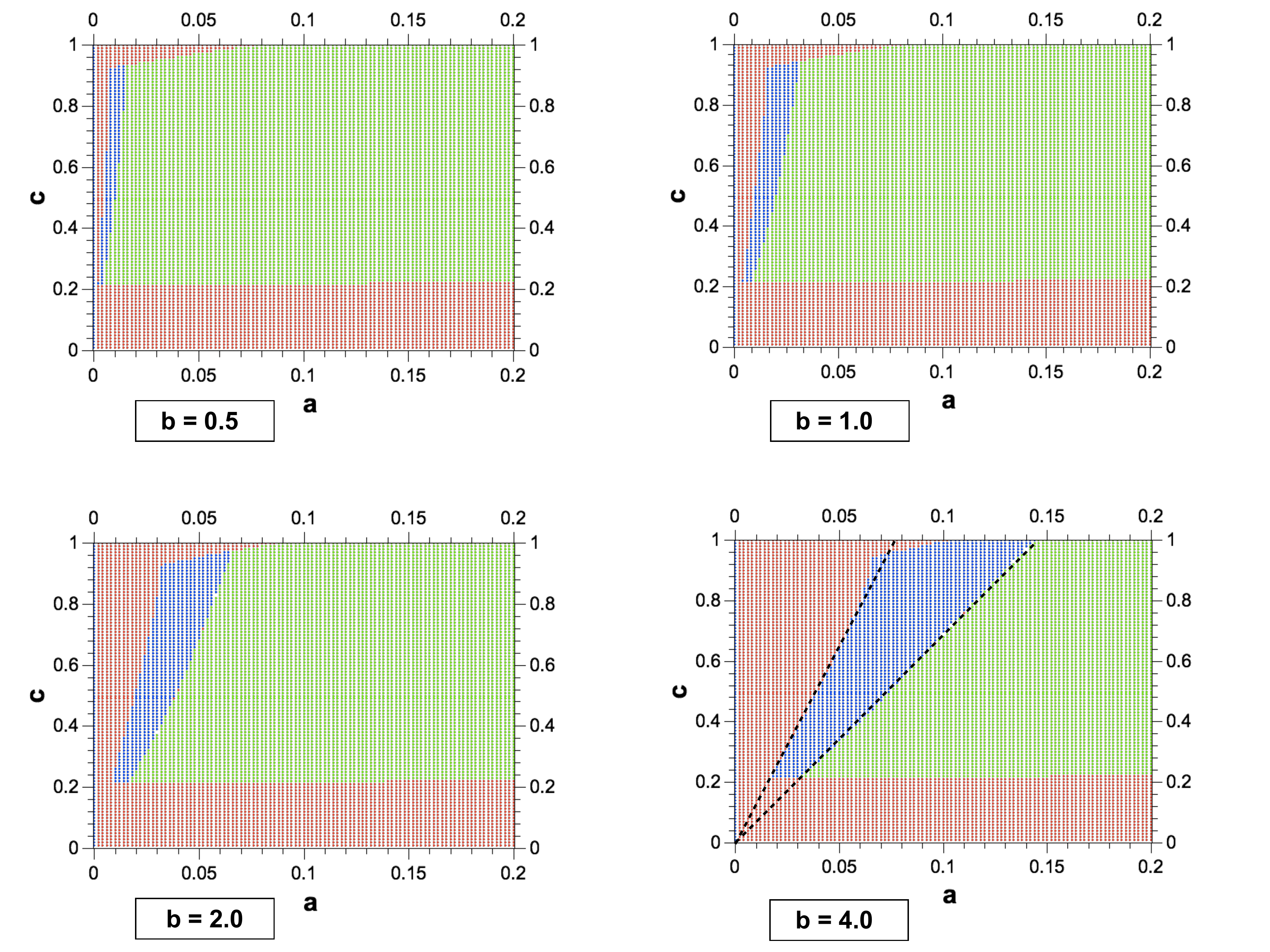}
\vspace{0.0cm}
\caption{Phase diagrams of flow regimes are depicted for some values of $b$. Simulations, which have explored the regions $0 \leq a \leq 0.2$ and $0 \leq c \leq 1$, have all been carried out for $\alpha = 0.2$ and $\alpha'=0$. Red, blue, and green points correspond to values of parameters associated, respectively, to particle-vortex decoupled flow regimes, backward-moving VRP states, and forward-moving VRP states. It turns out that backward-moving VRP states are approximately bounded from above and from below by physical lines, as particularly evidenced in the phase diagram for the case $b=4.0$ (straight dashed lines).}
\vspace{0.4cm}
\label{pd}
\end{figure}
\end{widetext}

(ii) No VRP states are found for $a$ small enough. Looking again at Eq. (\ref{udot}), we see that $a$ controls the viscous drag. As a consequence, the particle dynamics will be dominated by the superfluid pressure force at smaller values of $a$, if $b$ and $c$ are kept fixed. Due to inertia, however, the initial particle's backward motion will develop along evolution time scales which are much larger than the ones for the vortex dipole, which, then decouples from the particle. Its worth emphasizing that an analogous connection between small viscous dissipation and the suppression of particle trapping in quantum vortices has been previously noticed from three-dimensional numerical simulations of particle dynamics in vortex tangles \cite{kivo_serg_bar}.
\vspace{0.2cm}

(iii) For $a$ large enough, there is a transition, which is independent of $b$ and takes place at an approximately fixed value of $c$, between particle-vortex decoupled and forward-moving VRP flow regimes. We can understand this as follows. For large $a$, viscous drag is more relevant than the opposite pressure forces, so the particle moves along with the normal flow. As $c$ grows, the velocity field produced by the vortex dipole image (a fictious source inside the particle) gets more effective in advecting the vortex dipole back to the particle [see the discussion related to Fig. \ref{portrait}a)]. At some threshold value of $c$, determined essentially by Eqs. (\ref{zzpdot}) and (\ref{zzmdot}), the advection becomes, then, strong enough to stabilize the limit orbits of VRP states.
\vspace{0.2cm}

(iv) Relations (\ref{aa}) and (\ref{cc}) imply that 
\be
\frac{a}{c} \sim \frac{\rho_n \nu}{\rho_p \kappa} \ . \  \label{ac}
\ee
The constancy of $a/c$ against changes of $V$ or $R$ leads to interesting consequences. Assuming that $u$ is approximately constant at large asymptotic times, the time average of Eq. (\ref{udot}) gives, using (\ref{p_int}),
\be
u \simeq \frac{ a/c + ig}
 {a/c - ig}  \label{u_value} \ , \
\ee
where, working in the particle's reference frame ($z_p=0$),
\be
g = -\frac{ib}{\pi^2} \left \langle \frac{\hbox{Im}(z_+)\hbox{Re}(z_+)}{|z_+|^2}   \right \rangle \ . \
\ee
Eq. (\ref{u_value}) makes sense only for VRP states (otherwise, $u=1$, asymptotically). We expect, furthermore,
that at fixed $a/c$ and $b$, the above expectation value, associated to geometric features of the vortex dipole
stable orbits, do not change substantially as $a$ and $c$ are modified by identical factors. 
In this way, Eq. (\ref{u_value}) predicts that on sets of constant $a/c$ in phase-space slices -- we call them {\it{physical lines}} -- $u$ is approximately constant as well. Observed flow regimes for a given counterflow experiment are supposed to be related to the phases found along fixed physical lines. 
\vspace{0.2cm}

It turns out that phase-space regions of backward-moving VRP flow regimes are roughly bounded by physical lines. This is clearly indicated in Fig. \ref{pd} for the $b=4$ phase diagram. It is interesting to note that there are essentially no transitions between backward and forward-moving VRP flow regimes along physical lines. This is a puzzling fact, if we want to draw phenomenological lessons for three-dimensional superfluid flows, since such transitions have been observed in real experiments \cite{duda}.

A natural solution of this problem is to consider VRP states with vortex rings that carry additional circulation quanta. This amounts to replace the elementary circulation quantum $\kappa$ in Eq. (\ref{ac}) by $n \kappa$, where $n$ is a positive integer. The higher is $n$, the steeper will be the physical lines in phase-space slices as the ones portrayed in Fig. \ref{pd}. We conjecture, thus, that the observed particles that move along directions opposite to the normal flow (and are not trapped to vortex cores) can be described as 
backward-moving VRP states that contain topologically excited quantum vortex rings.

\begin{figure}[ht]
\hspace{0.0cm} \includegraphics[width=0.36\textwidth]{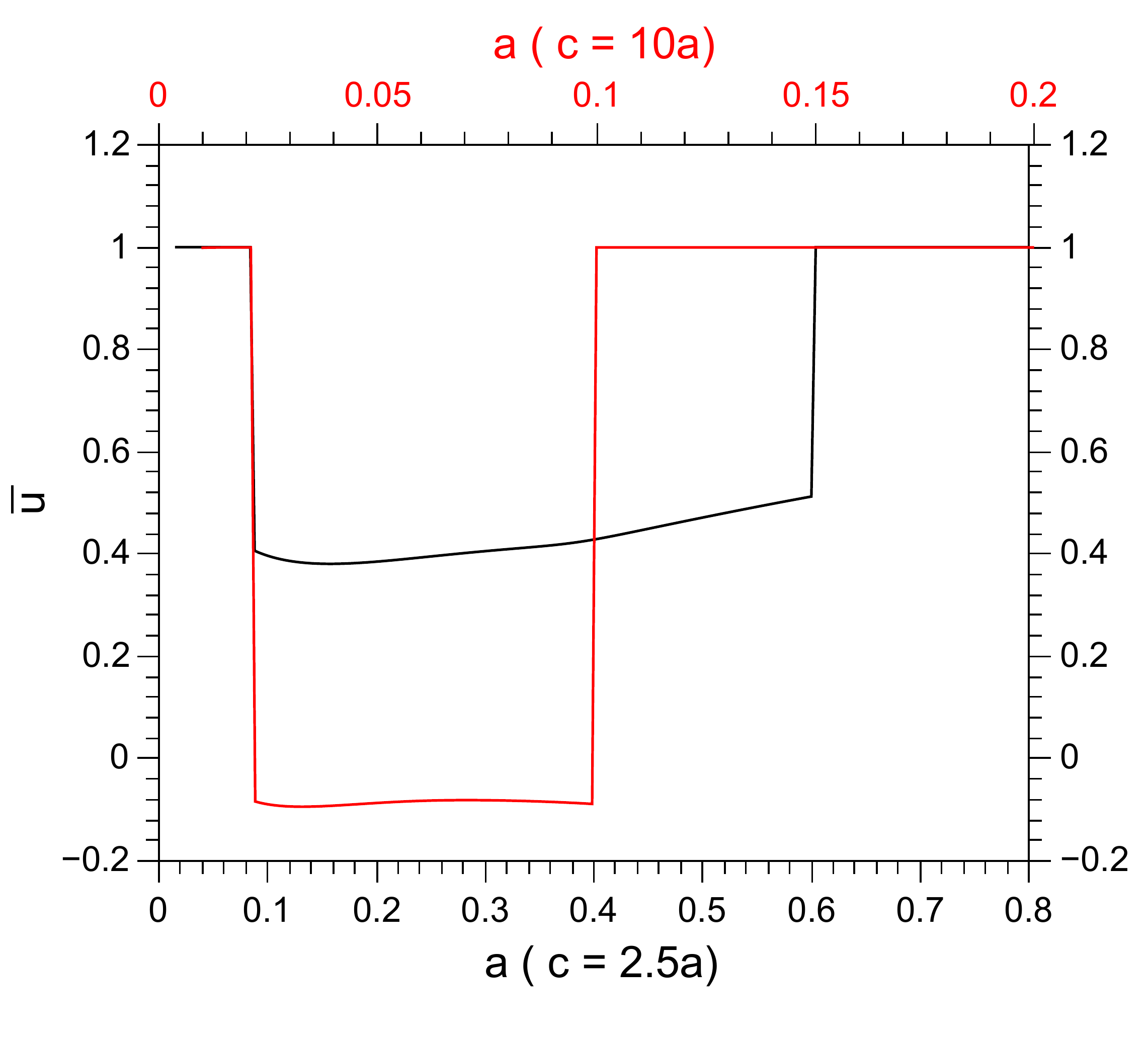}
\vspace{0.0cm}
\caption{Mean particle velocities $\bar u$ along the physical lines $c=2.5a$ (black line) and $c=10a$ (red line) as $a$ varies. They are associated to simulation parameters $\alpha=0.2$, $\alpha' =0$, and $b=4.0$.}
\label{pl}
\end{figure}

We have checked the role of physical lines for the case $b=4$, related to one of the phase diagrams showed in {\hbox{Fig. \ref{pd}}}. Results are given in {\hbox{Fig. \ref{pl}}}, for two distinct physical lines, $c=10a$ and $c=2.5a$, with the former crossing the phase-space region of backward-moving VRP states. These two examples support the approximate constancy of the asymptotic velocity of VRP states suggested by Eq. (\ref{u_value}).

These considerations offer a clear and simple explanation for the bimodal velocity distributions of tracking particles found in weak counterflow turbulence \cite{zhang_sci,mastracci_guo}, where the mean distance between quantum vortices is much larger than the typical particle's size, and one expects that VRP states should be rarer due to the attenuated rate of particle-vortex collisions. For stronger counterflow regimes, in contrast, the vortex tangles are denser, and VRP states should proliferate. As a result, the velocity distributions become unimodal, but peaked at ``anomalous" values.
\vspace{-0.0cm}

\section{Dynamical Balance of Stable Limit Orbits}

\begin{figure}[ht]
\hspace{0.0cm} \includegraphics[width=0.4\textwidth]{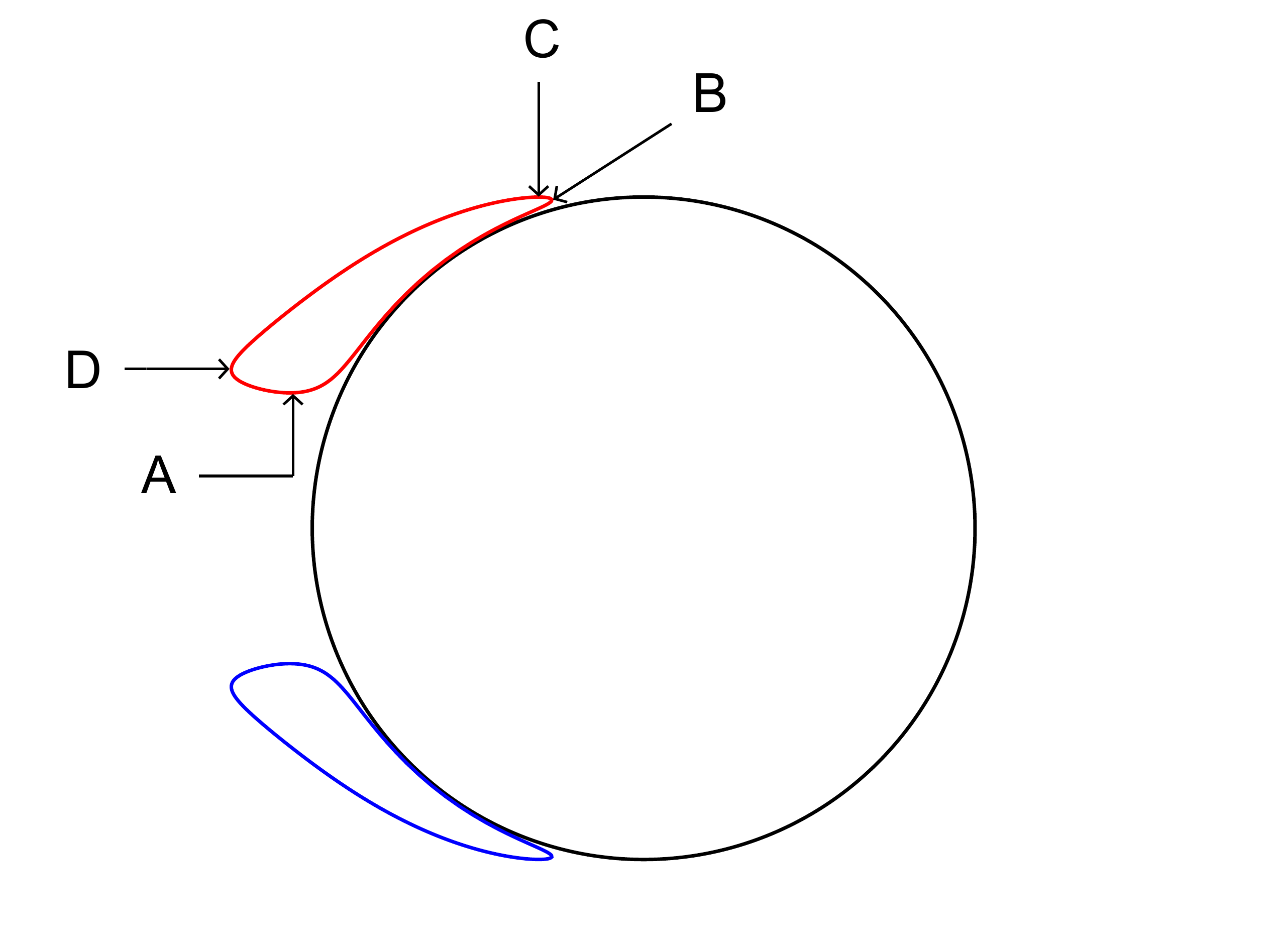}
\vspace{-0.5cm}
\caption{Returning points of vertical (A and C) and horizontal (B and D) vortex motions, as seen from the particle's reference frame. The positive (negative) vortex orbit, represented by the red (blue) line, is  counterclockwise (clockwise) oriented. Simulation parameters are $\alpha = 0.2$, $\alpha'=0$, $a=0.01$, $b=1.0$, and $c=0.5$.}
\label{tp}
\end{figure}

Picking up an arbitrary VRP state as a representative example, Fig. \ref{tp} shows its associated stable vortex orbits and returning points as well, as seen from the particle's reference frame.

\begin{figure}[b]
\vspace{-0.3cm}
~~\includegraphics[width=0.45\textwidth]{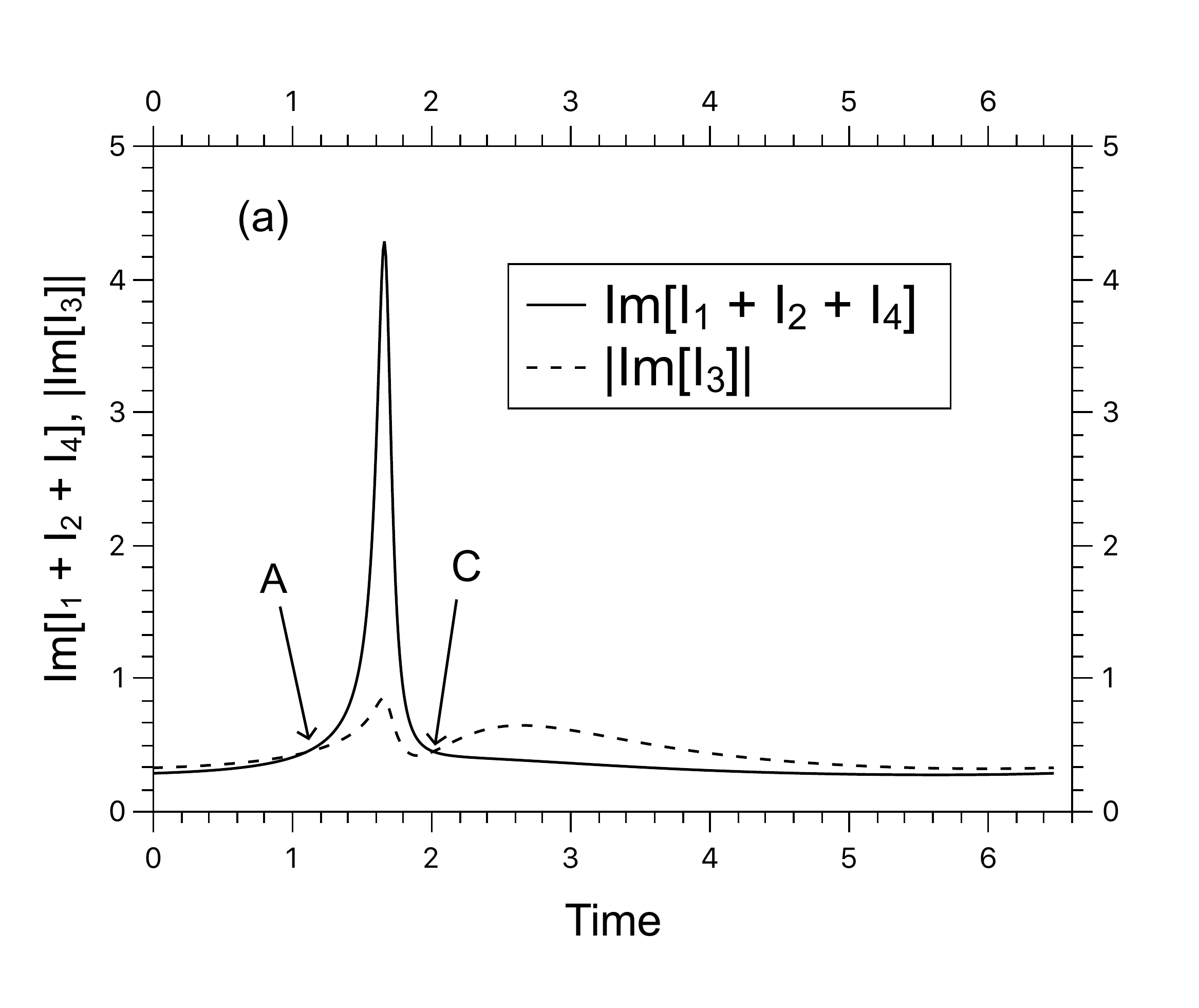}
\vspace{0.0cm}
\hspace{0.0cm} \includegraphics[width=0.45\textwidth]{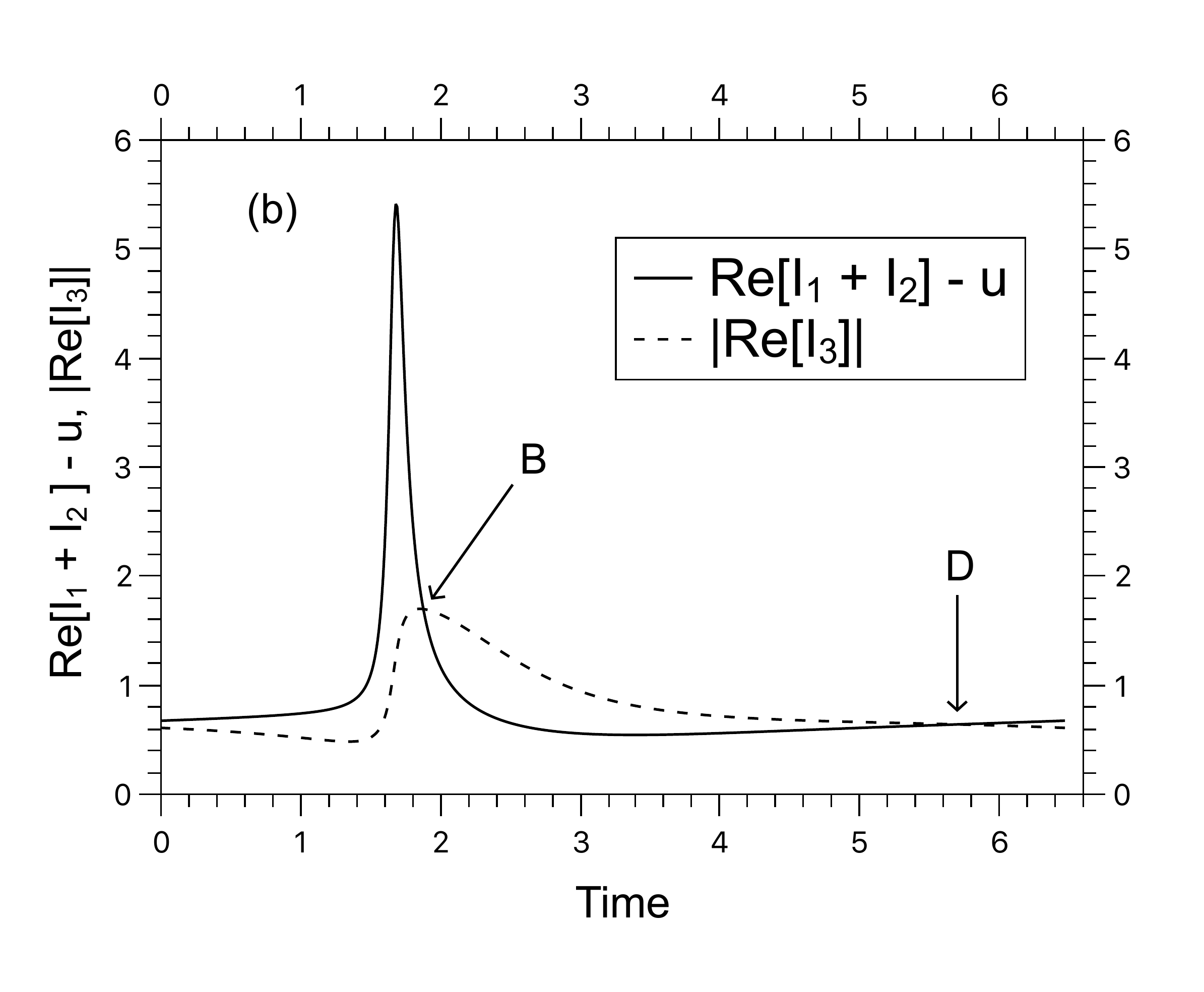}
\vspace{-0.5cm}
\caption{Dynamical balance of a stable limit orbit, as observed from the particle's reference frame, for the vertical (a) and horizontal (b) velocity contributions. The crossing points A, B, C, and D refer to the points indicated in {\hbox{Fig. \ref{tp}}}. Simulation parameters are $\alpha = 0.2$, $\alpha'=0$, $a=0.01$, $b=1.0$, and $c=0.5$.}
\label{DynBal}
\end{figure}
\vspace{0.0cm}

\begin{figure}[ht]
\hspace{-0.2cm} \includegraphics[width=0.48\textwidth]{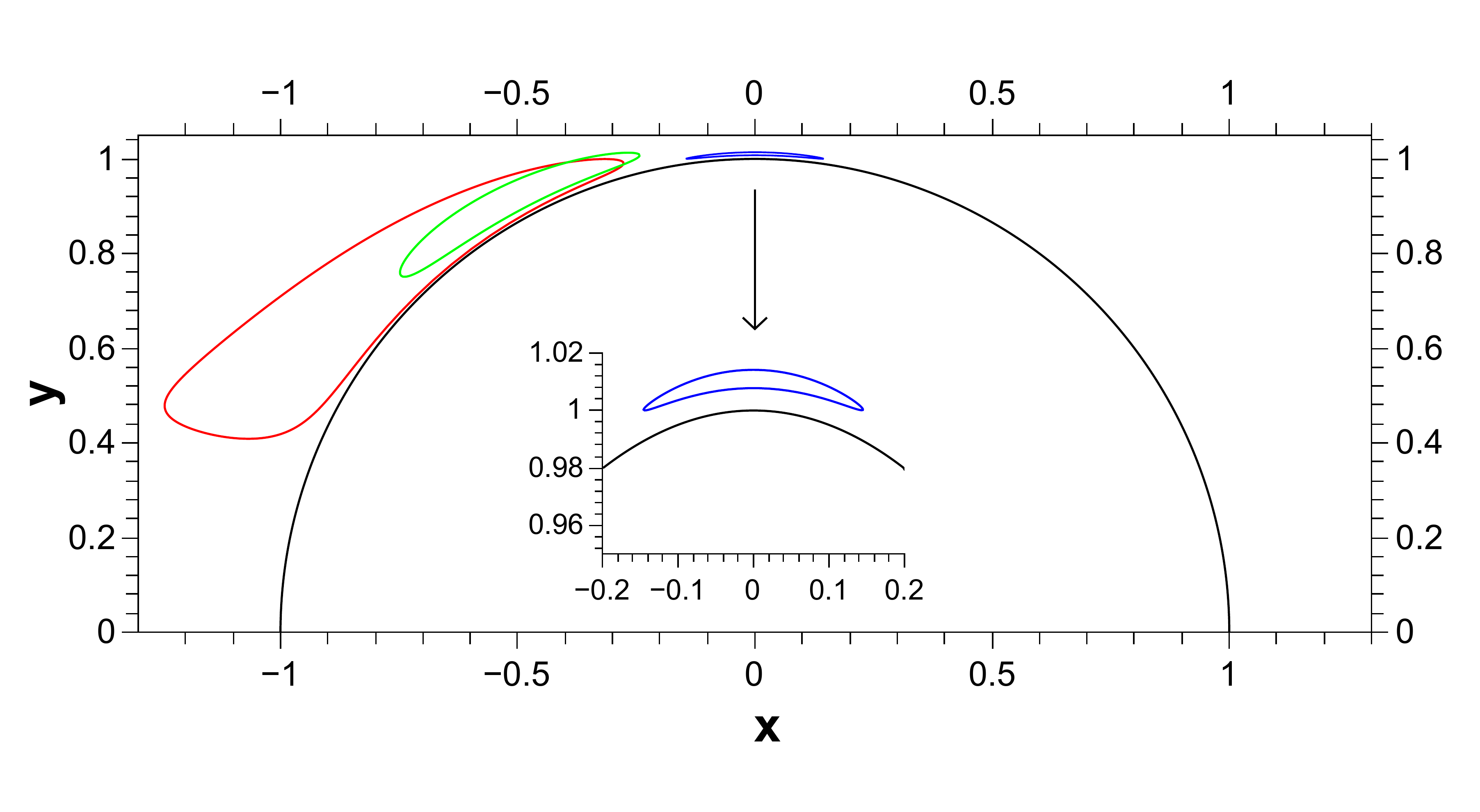}
\vspace{-0.5cm}
\caption{Stable cycles for $\alpha=0.0$ (blue), $0.1$ (green), and $0.2$ (red). In all the simulations, $\alpha'=0.0$, $a=0.01$, $b=1.0$ and $c=0.5$. The tiny orbit for the $\alpha=0.0$ is detailed in the inset. }
\label{alphas}
\end{figure}

Keeping Fig. \ref{tp} in mind, we now take a closer look on the various factors that contribute to the vortex dipole dynamics. For this purpose, we write the velocity of the positive vortex (in the ``laboratory" reference frame) as the sum of four independent contributions,
\be
\dot z_+ = I_1 +I_2 + I_3 + I_4  \ , \
\ee
where, according to (\ref{Wz}), (\ref{zzpdot}), (\ref{hatWp}), (\ref{Wp}), and (\ref{Wpart}),
\bea
I_1 &=& -i \frac{c}{2 \pi}\frac{z_+ - z_p}{|z_+-z_p|^2-1} \ , \ \label{I1} \\
I_2 &=& -i \frac{c}{2 \pi} \left( \frac{1}{z_+^* - z_-^*} 
 -\frac{z_- - z_p}{(z_+^*-z_p^*)(z_- - z_p)-1} \right ) \ , \ \nonumber \\ 
 \label{I2} \\
I_3 &=& -1 + \frac{1+u^*}{(z_+^*-z_p^*)^2} \ , \ \label{I3} \\
I_4 &=& \alpha'+i \alpha  \ , \  \label{I4}
\eea
are velocity contributions that come from the image of the positive vortex ($I_1$), from the negative vortex and its own image  ($I_2$), from the background superflow (modified by the presence of the particle) ($I_3$), and from the combination of Magnus and normal drag forces ($I_4$).

The imaginary and real parts of the above complex quantities (\ref{I1}-\ref{I4}) yield, respectively, contributions to the $y$ (vertical) and $x$ (horizontal) cartesian components of the positive vortex velocity vector. Graphs for convenient combinations of the $I's$ are provided in Fig. \ref{DynBal}, for complete cycles of vortex limit orbits. 

Taking into account that Re[$I_3$] and Im[$I_3$] are negative-valued functions of time, both Figs. \ref{DynBal}a and \ref{DynBal}b indicate that the returning points of vortex limit orbits are associated to situations where the background superflow (modified by the particle) cancels all the other combined velocity contributions due to the vortices and their images. This supports the stabilization picture of VRP states outlined in Sec. II (focused on the case of horizontal returning points).

The sharp peaks in the graphs of Figs. \ref{DynBal}a and \ref{DynBal}b are related to the pieces of vortex trajectories that are the closest to the particle's boundary, where vortices are advected by the velocity field induced by their images, which overcome the local superfluid background flow.

In our particular toy model, the Magnus contribution $I_4 = i \alpha$ for the positive vortex only affects its vertical velocity. The larger is $\alpha$, the broader and more stretched along the direction of the background superflow will be the vortex orbit, as Fig. \ref{alphas} shows. 

The more stretched vortex orbits associated to larger values of $\alpha$ are the ones which more effectively pull the particle along the superflow direction. We have verified, through further simulations, that the Magnus force is, in fact, a necessary phenomenological ingredient for the occurrence of backward-moving VRP states. It is important to stress, however, that for $\alpha$ large enough, the stability of either backward or forward-moving VRP states is lost, and we are left in this case with just vortex-particle decoupled flow regimes.

\section{Conclusions}

Dynamical bound states of vortex rings and tracking particles (VRP states) have been proposed as a way to understand puzzling PTV data obtained in counterflow turbulence. Our work, which focus on the essential dynamical aspects of VRP states, formulated with the help of simplified modeling tools, is closely related to the general proposal addressed in Ref. \cite{serg_bar_kivo}. We have taken advantage of the analytical structure of incompressible and inviscid two-dimensional fluid dynamics to discuss axisymmetric VRP states, likely to approximate the ones realized in real counterflow experiments. As discussed in detail by means of straightforward and fast numerical simulations, the present approach leads to several results that agree with previous observations and, furthermore, suggests the search of so far unsuspected phenomena, as the role of topologically excited vortex rings in backward-moving VRP states.

We call attention to the fact the VRP states produce pressure oscillations, due to the closed vortex ring orbits, with power spectrum peaked at frequencies of the order of $V/R$, which is estimated to be in the range of a few KHz for common counterflow experiments. It would be very interesting to look for such acoustic signals, in combination with PTV analyses.

Our toy-modeling approach, of course, is not aimed at providing accurate quantitative predictions. Rather, the idea is to point out interesting phenomenological aspects that could go unnoticed within more time-consuming three-dimensional simulations.


To conclude, we note that there is an appealing connection between the dynamics of VRP states and recent findings for the slow fall of seeds from wind-dispersed plants \cite{dandelion}. It has been found that the passive flight of dandelion seeds actually relies on the existence of approximate axisymmetric VRP states, with vertical symmetry axes. The vortex rings attached to dandelion seeds, however, do not oscillate and are created from a spontaneous flow separation mechanism, while the ones of superfluid VRP states are inherited from previously existing quantum vortex filaments.

\acknowledgments

This work was partially supported by CNPq.

\end{document}